\documentclass[fleqn,twoside,twocolumn,nofootinbib]{revtex4} 
\usepackage[sec]{ujp} 
\begin{document}
\title[CAPTURE OF ATOMS AND SMALL PARTICLES]
{CAPTURE~ OF~ ATOMS~ AND~ SMALL~ PARTICLES\\ IN AN OPTICAL TRAP
FORMED BY
SEQUENCES\\ OF COUNTER-PROPAGATING LIGHT PULSES\\ WITH A LARGE AREA}%
\author{V.I. Romanenko}
\affiliation{Institute of Physics, Nat. Acad. of Sci. of Ukraine}
\address{46, Nauky Ave., Kyiv 03028, Ukraine}
\email{vr@iop.kiev.ua}
\author{L.P. Yatsenko}%
\affiliation{Institute of Physics, Nat. Acad. of Sci. of Ukraine}%
\address{46, Nauky Ave., Kyiv 03028, Ukraine}%
\email{vr@iop.kiev.ua}%
\udk{535.372}  \razd{\seciv}

\setcounter{page}{893}%
\maketitle

\begin{abstract}
A new trap for atoms and small particles based on the interaction
between an atom and the field of counter-propagating light pulses that
are partially superposed in time has been proposed. A substantial
difference from the known analogs consists in that the atom--field
interaction is close to the adiabatic one, which allows a considerably
higher momentum to be transferred to the atom within the same time
interval and makes the trap smaller in size. It has been shown that,
owing to the dependence of the light pressure force on the atom
velocity, the atomic ensemble is cooled at its interaction with
the field.
\end{abstract}

\section{Introduction}

\label{introduction}Governing the atomic motions by applying light
fields has already gone beyond the scope of speculations, being
nowadays an ordinary means in physical experiments
\cite{Chu98,Coh98,Phi98,Bal00}. At the same time, new approaches in
this domain are proposed. For instance, it was shown
\cite{Rom00,Rom05,Dem06} that the variation of the momentum of an
atom at its interaction with counter-propagating light pulses, which
are partially superposed in time, can considerably exceed the double
momentum of a photon, which is a fundamental limit of momentum
transfer in the case where the atom interacts with
counter-propagating pulses one-by-one \cite{Kaz74,Neb74}.

As a rule, an atom in an optical trap permanently interacts with the field. This
can comprise an appreciable obstacle in physical experiments, in particular,
in laser spectroscopy. A possible way out from this situation is to organize
the interaction between the atom and the field in such a manner that the atom
would undergone the action of laser radiation only within a short time
interval, i.e. to construct light traps on the basis of counter-propagating
light pulses \cite{Fre95, Goe97, Bal05, Rom11}. It is important that, in the
cited works, the carrier frequencies of counter-propagating pulses are
identical, and, as a result, the force acting on the atom does not exceed
$2\hbar k/T$, where $T$ is the pulse-repetition period, and $\hbar k$ is the
photon momentum.

We suggest to combine the advantages suggested by a small perturbation
of an atom in the pulse trap with a capability to substantially
increase the light pressure force in the trap owing to the multiphoton
interaction. This combination forms a basis for the propositions of
how the momentum transferred to an atom can be made well above
$2\hbar k$ \cite{Rom00,Rom05,Dem06,Neg08}. With this goal in view,
the interaction between the atom and the field must be adiabatic, i.e.
the area of light pulses must considerably exceed $\pi$. The
difference between two schemes of the interaction between a two-level
atom and a field of counter-propagating pulses, which are examined
in works \cite{Rom00,Dem06,Neg08}, consists in the following: in
works \cite{Rom00,Neg08}, the atom interacts with the field of
counter-propagating pulses characterized by different carrier
frequencies, whereas, in work \cite{Dem06}, the current carrier
frequencies change linearly in time. As a result, the directions of
the momentum transferred to the atom during its interaction with the
field turn out different. In the former case, the direction of a variation of the atom
momentum is \textquotedblleft
counter-intuitive\textquotedblright, i.e. the momentum changes in
the direction of the momentum of the light pulse, which is the
second that interacts with the atom. In the latter case, it
coincides with the direction of propagation of the pulse, which is
the first that interacts with the atom. In Section 3, it is
demonstrated that, owing to this difference, the light pulses form a
potential barrier for atoms in the former case and a potential well
in the latter one. In Section 4, in order to determine whether those
barrier and well can be used to form an optical trap, we analyze the
dependence of the force that acts on the atom in the field of a pulse
sequence on the atom velocity and show that the force in the
potential well acts oppositely to the atom velocity and decelerates
the atom in the trap. However, in the case of light barrier, the
direction of the force acting on the atom depends on the sign of the
difference between the frequencies of the first and second light
pulses. As a result, the trap can be constructed in the bichromatic
field of pulses as well, provided that the barriers are spatially
located not far from each other.\looseness=1

\begin{figure}
\includegraphics[width=5cm]{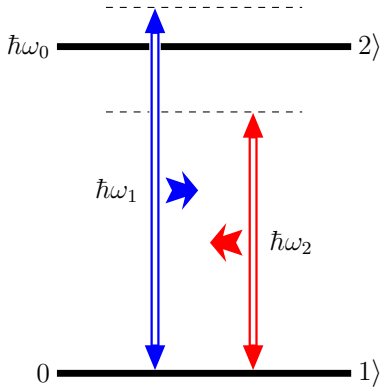}
\vskip-3mm\caption{Diagram of the interaction between an atom and
light pulses with the carrier frequencies $\omega_{1}$ and
$\omega_{2}$. The energy difference between the ground and excited
states equals $\hbar\omega_{0}$  }
\end{figure}

The proposed optical trap can also be applied to hold small particles
containing atoms with narrow spectral lines associated with the transition from
the ground into the excited state. The parameters of such a trap are evaluated
in Section~5. Short conclusions are made in Section~6.

\section{Basic Equations}

Consider a two-level atom with ground state $|1\rangle$, excited \ state
$|2\rangle$, and the frequency of the transition between them $\omega_{0}$. The
atom interacts with the field
\[\bm{\mathrm{E}}(t)=E_{1}(t)\bm{\mathrm{e}}
\cos[\omega_{1}t-k_{1}z+\varphi_{1}(t)]+\]
\begin{equation}
+E_{2}(t)\bm{\mathrm{e}} \cos[\omega_{2}t-k_{2}z+\varphi_{2}(t)]
\label{eq:E}%
\end{equation}
created by a sequence of counter-propagating pulses (see Fig.~1). Here,
$\omega_{1,2}$ are the carrier frequencies of pulses; $\varphi_{1,2}(t)$ the
corresponding phases, which are, generally speaking, dependent on time; $E_{1,2}(t)$ the
pulse envelopes; and $\mathbf{e}$ the unit vector of the pulse electric field
polarization. To make the notation simpler, the argument will not be
indicated below for the fields and the phases, as well as for the density matrix elements.

The interaction between the atom and the filed is described in the dipole
approximation. The Hamiltonian of this interaction looks like
\begin{equation}
H=\hbar\omega_{0}|2\rangle\langle2|-\bm{\mathrm{d}}_{12}|1\rangle
\langle2|\bm{\mathrm{E}}(t)-\bm{\mathrm{d}}_{21}|2\rangle\langle
1|\bm{\mathrm{E}}(t), \label{eq:Ham}%
\end{equation}
where $\mathbf{d}_{12}$ and $\mathbf{d}_{21}$ are the matrix elements of
the atomic dipole moment $\mathbf{d}$. Without any loss of generality
\cite{Sho90}, we may assume that $\mathbf{d}_{12}\mathbf{e}=\mathbf{d}%
_{21}\mathbf{e}$.

The equations for the occupation inversion $w=\varrho_{22}-\varrho_{11}$ and
for the coherence $\varrho_{12}$, where $\varrho_{nm}$ is the density matrix of
the atom, look as follows in the rotating-wave approximation \cite{Sho90}:
\[
\dot{w}=2\mathop{\mathrm{Im}}\varrho_{12}\left(\Omega_{1}e^{ikz-i\varphi_{1}-\frac{1}{2}i\delta{}t}+\mbox{}\right.\]
\[\left.\mbox{}+\Omega_{2}e^{-ikz-i\varphi_{2}+\frac{1}{2}i\delta{}t}\right)-\gamma\left(1+w\right),\]
\[\dot{\varrho}_{12}=-\frac{i}{2}\left(\Omega_{1}e^{-ikz+i\varphi_{1}+\frac{1}{2}i\delta{}t}+\mbox{}\right.\]
\begin{equation}
\left.\mbox{}+\Omega_{2}e^{ikz+i\varphi_{2}-\frac{1}{2}i\delta{}t}\right)w+\left(i\Delta-\frac{1}{2}\gamma\right)\varrho_{12},
\label{eq:dm}%
\end{equation}
where $\gamma$ is the inverse lifetime of the atom in the excited state,
$\Omega_{1}=-d_{12}E_{1}/\hbar$, $\Omega_{2}=-d_{12}E_{2}/\hbar$,
\begin{equation}
\delta=\omega_{1} -\omega_{2},\quad
\Delta=\omega_{0}-\frac{1}{2}\left(\omega_{1}+\omega_{2}\right).
\label{eq:d}
\end{equation}
Here, we selected the normalizing condition in the form $\varrho_{11}%
+\varrho_{22}=1$ and assume $k_1=k_2=k=\omega_0/c$.

We consider the interaction between an atom and a field created by two sequences
of counter-propagating light pulses with the repetition period $T$. One of the
sequences repeats the other with a definite time delay $t_{d}$ at the
atom location point:
\begin{equation}
\Omega_{1,2}=\Omega_{0}f(\eta_{1,2}),\label{eq:pulse}%
\end{equation}
where the function $f(\eta)$ describes the shape of a pulse envelope and has the
maximum value $f(0)=1$,
\begin{equation}
\eta_{1,2}=(2t\mp{}t_{d})/2\tau,\label{eq:eta}%
\end{equation}
$t_{d}$ is the difference between the arrival times of the maxima of pulses
propagating in the negative and positive directions along the $z$-axis at the
point, where the atom is located, and $\tau$ is the pulse duration.

While simulating the atom-field interaction, pulses of the Gaussian-like
shape are used, as a rule \cite{Ber98,Vit01}. It is known \cite{Rom06,Neg08}
that the function $\cos^{n}(\pi t/\tau)$ with the growing even power exponent
$n$ tends to $\exp(-t^{2}/\tau_{g}^{2})$, where $\tau_{g}=\tau\sqrt{2}/\left(
\pi\sqrt{n}\right)$, within the interval $\left\vert t\right\vert <\tau/2$.
For the numerical simulation, we selected the function $f(\eta)$ in the form
\begin{equation}
f(\eta)=\left\{
\begin{array}
[c]{ll}%
\cos^{4}(\pi{}\eta), & |\eta{}|<1/2\\
0, & |\eta{}|>1/2
\end{array}
.\right.  \label{eq:fcos}%
\end{equation}
In a vicinity of every pulse, this function is close to the Gaussian one
\begin{equation}
f_{G}(\eta)=\exp\left(  -2\pi^{2}\eta^{2}\right)  \label{eq:fG}%
\end{equation}
in the interval, where its value is not small (see Fig.~2). In comparison with
the Gaussian function (\ref{eq:fG}), function (\ref{eq:fcos}) selected by us for
the pulse simulation is, on the one hand, more convenient for numerical
calculations, because the Gaussian function has to be artificially cut off at
certain limits, and, on the other hand, it corresponds to real pulse envelopes
confined in time.

The area of a pulse, the envelope of which is described by function
(\ref{eq:fcos}), equals $\frac{3}{8}\Omega_{0}\tau$, amounting to
approximately 0.94 times the area of the Gaussian pulse close to it.

The pulse phases $\varphi_{1,2}$ depend quadratically on the time,
\begin{equation}
\varphi_{1,2}=\frac{\beta}{2}\eta_{1,2}^{2},\label{eq:phi}%
\end{equation}
so that the current frequency of each pulse changes linearly in time,
\begin{equation}
\varpi_{1,2}=\omega_{1,2}+\dot{\varphi}_{1,2}=\frac{\beta}{\tau^{2}}\left(
t\mp\frac{1}{2}t_{d}\right).\label{eq:dotphi}%
\end{equation}

The expression for the light pressure force acting on the atom \cite{Min86},
\begin{equation}
{\cal{F}}=2\frac{\partial \bm{\mathrm{E}}}{
\partial z}\mathop{\mathrm{Re}}
\varrho_{12}\bm{\mathrm{d}}_{21}\exp\left(i\left(\omega_{0}-\Delta\right)t\right),
\label{eq:F}%
\end{equation}
after its averaging over the period of oscillations with the frequency
$\omega$ for field (\ref{eq:E}) takes the form
\[
{\bar{\cal{F}}}=\hbar{}k\mathop{\mathrm{Im}}\varrho_{12}\left(\Omega_{1}e^{ikz-\frac{1}{2}i\delta{}t-i\varphi_{1}(t)}-\mbox{}\right.\]
\begin{equation}
\left.\mbox{}-\Omega_{2}e^{-ikz+\frac{1}{2}i\delta{}t-i\varphi_{2}(t)}\right).
\label{eq:FF}
\end{equation}
When the atom moves along the $z$-axis with the velocity $v$, its
coordinate changes together with the force acting on the atom.
Within the wavelength interval, the atom velocity is almost
constant. Therefore, in order to calculate the variation of the atom
momentum in time, we use the light pressure force averaged over the
wavelength $\lambda=2\pi c/\omega_{0}$, as it was done when
calculating the light pressure on atoms in a bichromatic field,
\begin{equation}
F=\dfrac{1}{\lambda}\int\limits_{z}^{z+\lambda}{\bar{\mathcal{F}}}(z^{\prime
})dz^{\prime}.\label{eq:Fav}%
\end{equation}

\begin{figure}
\includegraphics[width=\column]{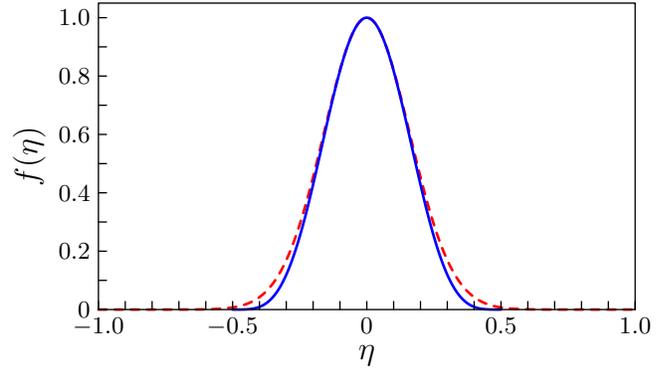}
\vskip-3mm\caption{Comparison between the function $f(x)$ describing
the envelope of light pulses (solid curve) and the most
approximating Gaussian function (dashed curve)  }
\end{figure}

\section{\textquotedblleft Heavy\textquotedblright\ Atom. Light Pressure Force
and Potential Energy}

For an atom moving along the $z$-axis with the velocity $v$, the detuning
$\delta$ of carrier frequencies of light pulses in time in the atom reference
frame changes according to the law
\begin{equation}
\delta=\delta_{0}-2k\int\limits_{0}^{t}v(t^{\prime})dt^{\prime},
\label{eq:delta}%
\end{equation}
where $\delta_{0}$ is the initial detuning.

Let us consider firstly the force of the light pressure on an atom in the
\textquotedblleft heavy\textquotedblright-atom approximation, when the
variation of the detuning of light pulse carrier frequencies from the transition
frequency $\omega_{0}$ during the atom motion weakly affects the magnitude of
light pressure force that acts on the atom. It is true, for instance, when the
second term in formula (\ref{eq:delta}) is small in comparison with the
first one. However, if we consider the pulses with current frequencies
$\varpi_{1,2}$ that vary in time, the role of $\delta_{0}$ is played by the
difference $\varpi_{1}-\varpi_{2}$, and this quantity must substantially
exceed $kv$ during the most part of the interaction time between the atom and the field.

\begin{figure}
\includegraphics[width=\column]{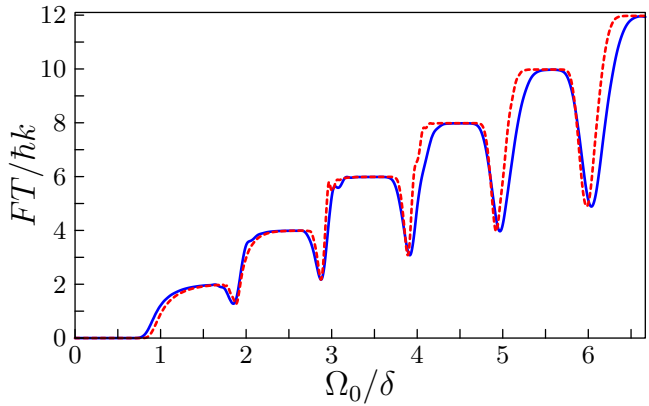}
\vskip-3mm\caption{Dependences of the light pressure force acting on the
atom in the field of counter-propagating pulses on the Rabi
frequency reckoned in $\delta$-units obtained by numerically solving
Eqs.~(\ref{eq:dm}) together with Eqs.~(\ref{eq:FF}) and
(\ref{eq:Fav}). The calculation parameters are $T=100\tau$,
$\beta=0$, $\Delta=0$, $t_{d}=0.25\tau$, $\gamma T=0.5$, and
$\delta\tau=300$ (solid curve) and 600 (dashed curve). The force was
averaged after the transient processes had terminated ($t>10T$)  }
\end{figure}

In Fig.~3, an example of the dependence of the light pressure force acting on the
atom on the Rabi frequency of light pulses is shown for the case of chirp-free
light pulses ($\beta=0$). In contrast to what was done in works
\cite{Rom00,Neg08}, we calculate the force acting on an atom during a
long $\;$($t\gg1/\gamma$) time of the interaction between the atom and the field,
rather than the average transferred momentum. Moreover, while calculating the
force acting on the atom, we used equations for the density matrix, which
allowed us, unlike the case with the Schr\"{o}dinger equation, to analyze the closed
scheme of the interaction between the atom and the field, when the atom, owing to
the spontaneous radiation emission, transits from the excited state in the
ground one.

As is seen from Fig.~3, the dependence of the light pressure force
acting on the atom on the Rabi frequency in the bichromatic field of
counter-propagating light pulses resembles stairs, with the latter
being observed more pronounced on the (dotted) curve that corresponds to
a larger area of light pulses. The altitude of
every stair is close to an even number in terms of $\hbar
k/T$-units. Note that the quantity $2\hbar k/T$ is the maximum
force of the light pressure on the atom, which can be obtained in the
field of light pulses, which alternatively interact with the atom.
Really, when the atom absorbs a photon from a light pulse, its
momentum can change to that of the photon, $\hbar k$, and when it
interacts with the counter-propagating light pulse, its momentum can
change again by $\hbar k$ in the same direction.\looseness=1

The behavior of the dependence shown in Fig.~3 originates from the
adiabaticity of the interaction between the atom and the bichromatic field
\cite{Rom00,Neg08}. In such a field, the atom is characterized by a spectrum
of quasi-energies \cite{Zel66} with the period $\hbar\delta$, which
corresponds to the momentum change by $2\hbar k$. When the atom adiabatically
interacts with the field, it occupies one of its characteristic states, and
the Landau--Zener transitions into other states with close quasi-energy values
become probable. As a result, the momentum of the atom during its interaction
with a pair of pulses changes by a value multiple of $2\hbar k$. The
coefficient of multiplicity is close to $\Omega_{0}/\delta$ (for more details,
see works \cite{Rom00,Neg08}).

If the atom moves along the $z$-axis, the delay $t_{d}$ between pulses changes
in the reference frame connected with the atom. For instance, if the atom is
located at a point, at which light pulses arrive simultaneously (let it be
selected for the coordinate origin), $t_{d}=0$. It is easy to see that, for
the atom with the coordinate $z$, the time delay $t_{d}$ at this point between the arrivals
of pulses that propagate in the positive and negative directions
of the $z$-axis is equal to $2z/c$.

Figure~4,$a$ illustrates an example of the dependence of the force
that acts on the atom on the atom coordinate obtained for the case
of a field with constant current frequencies ($\beta=0$). One can
see that the light pressure force vanishes in the region, where
light pulses alternatively interact with the atom. This circumstance
is associated with the fact that, owing to a large area of pulses
and a large detuning of light pulse carrier frequencies from the
resonance one, i.e. $\left\vert \omega_{1}-\omega_{0}\right\vert
=\left\vert \omega_{2}-\omega_{0}\right\vert =\delta/2\gg1/\tau$,
the atom is permanently in one of its adiabatic states. As a result,
the coherent population return takes place \cite{Vit01-117}; the
atom states at the beginning and the end of the interaction of the
atom with the field coincide; therefore, the atom momentum does not
change, if the atom interacts with a \textit{single} light pulse, so
that the light pressure force equals zero.

In Fig.~4,$b$, the coordinate dependence of the atomic potential energy $U$
determined by the equation
\begin{equation}
F=-\frac{dU}{dz}\label{eq:U}%
\end{equation}
is depicted. The reference mark for the potential energy was so selected that
it equals zero beyond the pulse superposition region. As one can see, the
counter-propagating light pulses form a potential barrier for atoms.

\begin{figure}
\includegraphics[width=\column]{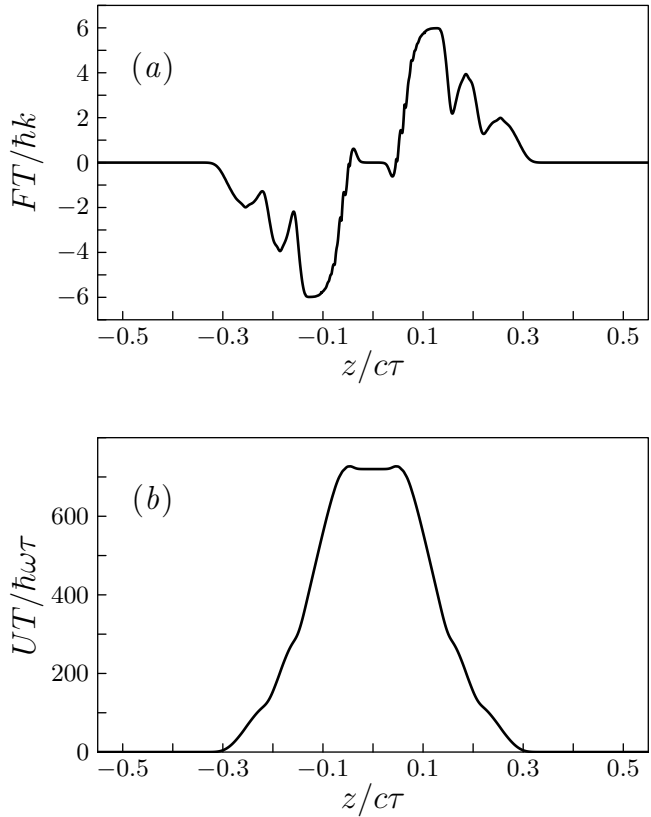}
\vskip-3mm\caption{Dependences of the light pressure force acting on
the atom in the field of counter-propagating pulses ($a$) and the
atom potential energy on the atom coordinate ($b$) determined by
numerically solving Eqs.~(\ref{eq:dm}) together with
Eqs.~(\ref{eq:FF}) and (\ref{eq:Fav}). The calculation parameters
are $T=100\tau$, $\beta=0$, $\Delta=0$, $\gamma T=0.5$,
$\delta\tau=300$, and $\Omega_{0}\tau=1000$. The force was averaged
after the transient processes had terminated ($t>10T$)  }
\end{figure}

In Fig.~5,$a$, an example of the coordinate dependence of the force acting on
the atom is shown for the case of a field with current frequencies linearly
varying in time ($\beta=200$). In contrast to Fig.~4,$a$, the light pressure
force differs from zero, being close to $2\hbar k/T$, in the region, where
light pulses interact, in turn, with the atom. This circumstance is associated
with an adiabatically quick passage of the resonance \cite{Sho90}, when either of
light pulses interacts with the atom, which results in that the atom transits
from the ground state into the excited one, which is accompanied by the
absorption of a photon and a momentum change by $\hbar k$. When the atom
interacts with the counter-propagating pulse, the induced photon emission takes
place, and the atom changes its momentum by another $\hbar k$ in the same
direction \cite{Neb74}.

\begin{figure}
\includegraphics[width=\column]{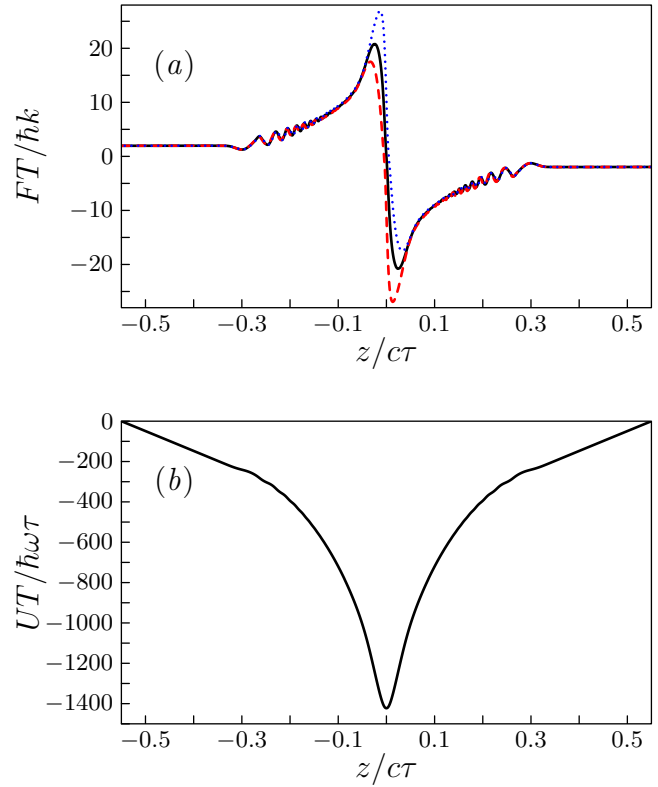}
\vskip-3mm\caption{Dependences of the light pressure force acting on
the atom in the field of counter-propagating pulses ($a$) and the
atom potential energy on the atom coordinate ($b$) determined by
numerically solving Eqs.~(\ref{eq:dm}) together with
Eqs.~(\ref{eq:FF}) and (\ref{eq:Fav}). The calculation parameters
are $T=100\tau$, $\beta=200$, $\Delta=0$, $\gamma T=0.5$,
$\Omega_{0}\tau=1000$, and $\delta\tau=0$ (solid curve), $-2$
(dashed curve), and 2 (dotted curve). The force was averaged after
the transient processes had terminated ($t>10T$)  }
\end{figure}

In Fig.~4,$b$, the coordinate dependence of the atomic potential energy is
depicted for a field with current frequencies that vary linearly in time. In
contrast to the interaction between the atom and the field created by
counter-propagating pulses with fixed current frequencies, the potential well
for the atom is formed in this case.

Hence, the field created by counter-propagating light pulses with fixed
frequencies or frequencies that linearly vary in time allows potential
barriers or potential wells to be created in the space, which can be used to
control the motion of atoms or nanoparticles containing atoms with a narrow
absorption line or to localize them in a definite space region. For instance,
the potential well can be used directly to hold atoms in a small spatial
region. At the same time, two potential barriers, which are simple to be
created with the use of two pairs of counter-propagating pulse sequences, can be
used to form an optical trap between them.

While creating traps, the important issue is the dependence of the light pressure
force on the atom velocity. It can result in a growth of the atom energy followed
by a probable escape of the atom beyond the trap boundaries, as well as in its
reduction, which corresponds to a stable atom localization in the trap.

\section{Optical Trap Formed by Sequences of Counter-Propagating Pulses}

In the course of long-term interaction between the atom and the field, the velocity
of the former changes. Simultaneously, according to Eq.~(\ref{eq:delta}), the
difference between the carrier frequencies of light pulses that act on the
atom also changes in the atom reference frame. Let us consider firstly light
pulses with a fixed carrier frequency ($\beta=0$). Let $\delta_{0}>0$ and
$v>0$. Then, if $v$ grows, $\delta$ decreases, and $\Omega_{0}/\delta$
increases. At the same time, one can see from Fig.~2 that, in general, the
light pressure force that acts on the atom grows with $\Omega_{0}/\delta.$
Hence, if $v$ is positive and grows, the force acting on the atom
also grows. However, if $v<0$ and $\left\vert v\right\vert $ grows, similar
speculations bring us to a conclusion that the light pressure force decreases.

\begin{figure}
\includegraphics[width=\column]{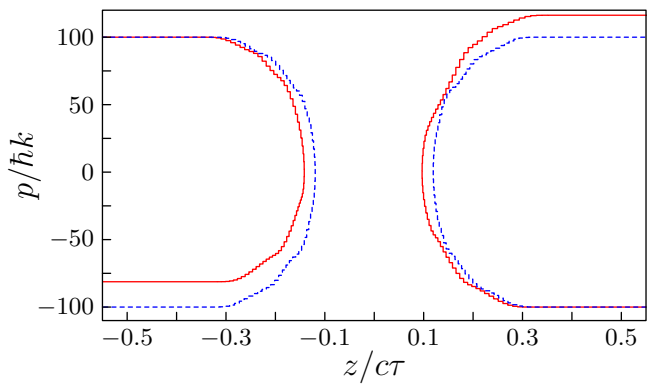}
\vskip-3mm\caption{Phase trajectories of an atom in the field of
sequences of counter-propagating pulses. The calculation parameters
are $T=100\tau$, $\beta=0$, $\Delta=0$, $\gamma T=0.5$,
$\delta\tau=300$, $\Omega_{0}\tau =1000$, $\alpha=10^{-6}$, and
$\mu=1$ (solid curves) and 0 (dashed curves). The force was averaged
after the transient processes had terminated ($t>10T$). The
left-hand side of the figure corresponds to the initial momentum
$p_{\text{ini}}=100\hbar k$ and the right-hand side to $p_{\text{ini}%
}=-100\hbar k$  }
\end{figure}

Let the atom move along the $z$-axis toward the barrier created by
the sequences of counter-propagating pulses. Let a pulse with a
higher carrier frequency (i.e. $\delta>0$) act on it first. When
having achieved the barrier, the atom is decelerated by the light
pressure force, and its velocity diminishes. After the atom having
been reflected from the barrier, the atom velocity grows. However,
since it is directed oppositely, the light pressure force,
according to the reasoning given above, is less than that when the
atom moves toward the barrier. Therefore, the atom, after having
been reflected from the barrier located at larger $z$'s, flies
away from it with a velocity lower than the initial one, if
$\delta>0$. However, if the atom collides with the barrier when
moving to the negative direction along the $z$-axis, then,
according to similar speculations, its velocity after the
reflection from the barrier will exceed the initial one. In
Fig.~6, the phase trajectories of the atom are shown for both
cases of motion, namely, with the initial velocity directed in the
positive (the right part of the figure) and negative (the left
part of the figure) directions of the $z$-axis. The trajectories
(solid curves) illustrate an increase or a reduction of the atom
velocity, when the atom is reflected from the barrier, depending
on the direction of its initial velocity. The dotted curve denotes
the phase trajectory obtained in the approximation of a very heavy
atom, i.e. when $kv\lll\delta_{0}$.

While calculating the atom momentum change, we neglected the influence of a variation of the atom
velocity on the light pressure force in the course of interaction
between the atom and the field created by a pair of pulses. The change of the atom
momentum was calculated as follows. First we calculated the variation of the atom
momentum,
\begin{equation}
\Delta{}p=\int\limits_{t_{\text{ini}}}^{t_{\text{fin}}}F\,dt,\label{eq:dP}%
\end{equation}
where $t_{\text{ini}}$ and $t_{\text{fin}}$ are the initial and final times of
the interaction between the atom and the field of a pulse pair. Then, we
calculated the variation of the atom velocity,
\begin{equation}
\Delta{}v=\frac{\Delta{}p}{M},\label{eq:dv}%
\end{equation}
where $M$ is the atom mass. The obtained $v$-value and the corresponding
$\delta$-value calculated according to Eq.~(\ref{eq:delta}) were used to
repeat the calculation procedure for the next pair of pulses. For
illustrative calculation, we selected the overestimated values of parameters
$\mu=\hbar k^{2}\tau/M$ and $\alpha=\hbar k/cM$, which govern the dependence
of the dimensionless detuning $\delta\tau$ on the atom momentum in $\hbar
k$-units,
\begin{equation}
\delta\tau=\delta_{0}\tau-\mu\,\left(  \frac{p}{\hbar{}k}\right)
\label{eq:deltap}%
\end{equation}
and the dependence of the dimensionless coordinate $\Delta z/c\tau$ on the
dimensionless time $\Delta t/\tau,$
\begin{equation}
\frac{\Delta{}z}{c\tau}=\alpha\left(  \frac{p}{\hbar{}k}\right)  \frac
{\Delta{}t}{\tau},\label{eq:z}%
\end{equation}
respectively.

By selecting the difference between the frequencies of light pulses in such a
way that the velocity of an atom after its interaction with the barrier formed by
those pulses would decrease and placing two barriers beside each other, it is
possible to obtain a one-dimensional trap for atoms.

Now, let us analyze a possibility of creating a trap on the basis of
counter-propagating pulses with varying current frequencies. In Fig.~5,$a$,
the dependences of the force on the atom coordinate are plotted for $\delta$'s,
which are identical by absolute value, but different by sign. The positive
(negative) sign of $\delta$ at $\delta_{0}=0$\ corresponds, according to
Eq.~(\ref{eq:delta}), to the motion of the atom in the negative (positive) direction
of the $z$-axis. It is easy to see that the light pressure force is less, when the
atom moves away from the coordinate origin than when it moves in the opposite
direction. Therefore, one may expect a stable holding of the atom by the field of
counter-propagating pulses with varying frequencies.

Figure~7 illustrates the motion of an atom in a trap created by counter-propagating
sequences of light pulses with frequencies varying in time. One can see that,
owing to the Doppler shift of carrier frequencies for light pulses in the atom
reference frame, the atom becomes ultimately localized at the trap center. It
should be noted that this fact does not mean that all atoms will gather at the
center, because the applied model of classical motion of an atom in the light field
does not take the momentum diffusion into account \cite{Min86}. In order to
evaluate the localization region, which constitutes a subject of our
subsequent researches, the quantum-mechanical description of the motion of atoms
should be applied.

Unlike the light trap formed by sequences of counter-propagating $\pi$-pulses
or pulses with a small area \cite{Fre95,Goe97,Bal05,Rom11}, the light pressure
force in the region, where pulses are spatially superposed, does not diminish,
but grows. As a result, the potential well in this region becomes much deeper.

Depending on the way used to create a trap, i.e. whether pulses with a fixed
frequency or a current frequency that varies in time are applied, the pulse
intensity can differ considerably. In work \cite{Dem06}, the intensities required
to transfer the same momentum to a helium or rubidium atom in the cases
where the nanosecond pulses acting on the atom have either a fixed carrier
frequency or a current carrier frequency that varies in time, were compared.
According to the estimations given in that work, the required intensity
amounts to about 1~MW/cm$^{\mathrm{2}}$ in the former and to a few
kW/cm$^{\mathrm{2}}$ in the latter case.

\section{Traps for Nanoparticles}

The optical trap proposed for atoms can also be used for capturing and holding
nanoparticles. For this purpose, the particles must contain \textquotedblleft
active\textquotedblright\ atoms with narrow absorption lines, and the
concentration of such atoms in the particles must be sufficiently high. Let us
evaluate the minimum required concentration of \textquotedblleft
active\textquotedblright\ atoms in a nanoparticle. The evident requirement for
the holding and manipulation of a nanoparticle to be possible is a substantial
predominance of the light pressure force, $F$, over the gravitation one, $Mg$.
Let us evaluate the light pressure force on an atom. In the region of
spatial pulse superposition, it is of the order of $10\hbar k/T$. For the
wavelength $\lambda\approx600$\textrm{~nm}, the pulse repetition frequency
$T^{-1}\approx100~\mathrm{MHz}$, and an atom with $M\approx50~\mathrm{amu,}$
we obtain $F/mg\approx10^{6}$. Hence, even if the concentration of
\textquotedblleft active\textquotedblright\ atoms in nanoparticles exceeds
0.001\%, the nanoparticles can be held in optical traps formed by
counter-propagating light pulses with a large area.

\begin{figure}
\includegraphics[width=\column]{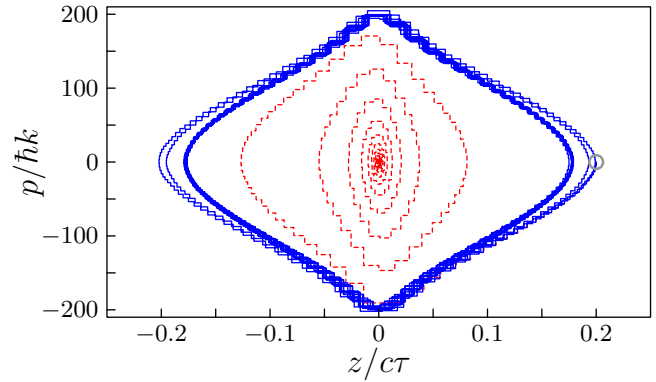}
\vskip-3mm\caption{Phase trajectories of an atom in the field of
sequences of counter-propagating pulses. The calculation parameters
are $T=100\tau$, $\beta=200$, $\Delta=0$, $\gamma T=0.5$,
$\delta=300$, $\Omega_{0}\tau=1000$, $\alpha=10^{-6}$, and
$\mu=0.01$ (dashed curve) and 0 (solid curve). The force was
averaged after the transient processes had terminated ($t>10T$). The
initial coordinate $z_{\text{ini}}=0.2c\tau$ is denoted by a gray
ring on the right  }
\end{figure}

\section{Conclusions}

We have proposed optical traps, which are based on the adiabatic interaction
between atoms, including atoms in nanoparticles, and sequences of
counter-propagating light pulses with a large area. In comparison with traps
on the basis of $\pi$-pulses and pulses with a small area, the new traps are
characterized by a larger light pressure force in the region of spatial pulse
superposition, which allows atoms and nanoparticles to be localized in a
smaller volume. In addition, owing to the dependence of the light pressure force in
those traps on the atom velocity, the energy of atoms in the trap decreases.
However, the issue concerning the maximum cooling of atoms in the traps remains
opened, because the classical description of the motion of an atom in the trap, which
was used in this work, is not enough for its study.

\vskip3mm The work was executed in the framework of the State
goal-oriented scientific and engineering program \textquotedblleft
Nanotechnologies and Nanomaterials (2010-2014)\textquotedblright\
(themes 1.1.4.13 and 3.5.1.24).

\rezume{%
УТРИМАННЯ ~АТОМІВ~ І~ МАЛИХ~ ЧАСТИНОК \\ОПТИЧНОЮ~~ ПАСТКОЮ, ~~СФОРМОВАНОЮ\\
ПОСЛІДОВНОСТЯМИ ~~ЗУСТРІЧНИХ \\СВІТЛОВИХ ІМПУЛЬСІВ ВЕЛИКОЇ
\\ПЛОЩІ}{В.І. Романенко, Л.П. Яценко} {Запропоновано нову пастку для
атомів і малих частинок, в основі якої -- взаємодія атома з полем
зустрічних імпульсів, що частково накладаються у часі. Суттєвою
відмінністю від відомих аналогів є близька до адіабатичної взаємодія
атома з полем, що дозволяє протягом того ж часу взаємодії передати
атому значно більший імпульс і зменшити розмір пастки. Показано, що
завдяки залежності світлового тиску  від швидкості під час взаємодії
з полем відбувається охолодження ансамблю атомів.}

\end{document}